\documentclass[prb,showpacs,reprint,amsmath,amssymb,floatfix,superscriptaddress,nobibnotes]{revtex4-1}

\usepackage{graphicx}
\usepackage{amsmath} %American mathematical society
\usepackage{color}
\usepackage{boxhandler}% alligns captions to the left
\usepackage[bbgreekl]{mathbbol}
\usepackage{multirow}%creats multirow in the Table
\usepackage{diagbox}%creats a diagonal box in the Table.1 to distinguish two modes
\allowdisplaybreaks[3]

\newcommand{\mh}{\mathbf{h}}

\DeclareSymbolFontAlphabet{\mathbbl}{bbold}

\begin{document}
\title{Charge-pseudospin coupled diffusion in semi-Dirac graphene: pseudospin assisted valley transport}
\author{Saber Rostamzadeh}	
\thanks{saber.rostamzadeh@universite-paris-saclay.fr}
\affiliation{Department of Physics, Istanbul University, Vezneciler 34134, Istanbul, Turkey}
\affiliation{Laboratoire de Physique des Solides, Universit\'e Paris Saclay,
CNRS UMR 8502, F-91405 Orsay Cedex, France}

\author{Mustafa Sarisaman}
\affiliation{Department of Physics, Istanbul University, Vezneciler 34134, Istanbul, Turkey}
%\date{}
\begin{abstract}
Modifying the hexagonal lattices of graphene enables the repositioning and merging of the Dirac cones which proves to be a key element in the use of these materials for alternative electronic applications such as valleytronics. Here we study the nonequilibrium transport of carriers within a system containing two Dirac cones in both standard graphene and semi-Dirac graphene. In the latter, the lattice modifications cause the relativistic and parabolic dispersion bands to coexist, furnishing the Fermi surface with a rich pseudospin texture and a versatile Dirac cones separation. We construct a kinetic theory to investigate the carrier diffusion and uncover that the pseudospin index contributes to the particle current and, like the real spin, can induce a magnetoelectric effect, and argue that the pseudospin-charge coupling can be utilized to design a pseudospin filter. We explore the charge dynamics inside a quasi-one-dimensional conductor using the drift-diffusion model and detect the pseudospin accumulation at the sample boundaries. We find that, while, for graphene, the accumulation contributes to an extra voltage drop between the sample interfaces, the semi-Dirac system presents a similar accumulation that is strikingly equipped with valley polarization, signifying an essential tool for the control of valley manipulation and chirality transport using the pseudospin.
\end{abstract}	
\maketitle

\section{Introduction}

The electronic bands in graphene coalesce in the Brillouin zone near two distinct momenta, called valleys\cite{PhysRevLett.99.236809}, in which the carriers exhibit the Dirac-like linear dispersion relation. Dirac valleys separate with a large momentum such that the valleys intermix when sharp impurities and point scatterings are available. The conic Dirac spectrum is associated with these points and is generally characterized by two auxiliary isospin indices: the (momentum) valley and the (sublattice ) pseudospin indices. From a fundamental perspective, graphene is an excellent toy model for studying quantum transport in mesoscopic settings. This quality is due to the further degrees of freedom available for the carriers offered by the distinct low-energy electronic band structure. The extra degrees of freedom provide a new paradigm for carrier transport, in which the information transport occurs not by the charge but via the additional isospin indices. Thus a potential element for the novel electronic industry such as pseudospintronics,\cite{pesin2012spintronics,san2009pseudospin,PhysRevB.77.041407} and valleytronics\cite{schaibley2016valleytronics,yao2008valley,schomerus2010helical}, just like the rise of spintronics utilizing the spin index  \cite{han2014graphene,RevModPhys.92.021003,choudhuri2019recent}.

Solid progress has been made in harnessing the valley quantum index in graphene using different extrinsic methods such as circular light polarization, triangular wrapping and Fermi surface distortion, external gauge fields, strain, etc.,\cite{gorbachev2014detecting,behnia2012polarized,jiang2013generation,PhysRevLett.100.236801,pereira2009strain,low2010strain}. On implementing the valley index in low-bias transport, graphene nanoconstrictions with definite boundaries prove to work as a valley polarizer\cite{rycerz2007valley,recher2010quantum}. There are still obstacles to integrating these structures into electronic devices, such as controllably breaking the valley degeneracy and sustaining long valley polarization and electric manipulation\cite{vitale2018valleytronics,lins2020perfect}. This motivates the search for new engineered honeycomb lattice materials with better valleytronics functionality. 

Compared to the valley, the pseudospin index is more elusive to probe due to its eccentric behaviour to the external stimulations and the inextricable nature of this degree of freedom. It is, however, established that under conditions, the pseudospin in graphene can induce ferromagnetic order \cite{majidi2011pseudospin,macdonald2012pseudospin,schomerus2010helical,trushin2015ultrafast,PhysRevB.77.041407,san2009pseudospin}. Besides standard graphene, in an engineered honeycomb lattice, and materials with modified Dirac dispersion\cite{goerbig2008tilted,PhysRevB.100.075438} a sizable pseudospin polarization is observed\cite{jung2020black}. Strain modification of the graphene lattice also redistributes the charge density of the two sublattices and leads to pseudospin polarization\cite{georgi2017tuning}. In photonic graphene, the pseudospin pertains to an angular momentum that can interact with optical beams inducing vortex-generation\cite{song2015unveiling}. In the Kekule distorted graphene, adjusting the intervalley distance via a parameter locks the valley degrees of freedom to the pseudospin and hence to the direction of momentum \cite{andrade2019valley,gamayun2018valley}; moreover, valley splitting and polarization are observed in the deformed graphene lattice \cite{stegmann2018current,lantagne2019dispersive,aktor2019topological}. These studies demonstrate that engineering the hexagonal lattice can turn the elusive pseudospin into a functional, practical element, just like the valley index with similar potential for electronic applications.

Although much effort was put into analysing the valley and pseudospin indices separately for their latent electronic abilities, a natural question arises: whether their coupled dynamics could prove more useful.
In this paper, we address this question by considering a modified graphene lattice where the two degrees of freedom can be jointly inspected and examine pseudospin-valley dynamics from a diffusive point of view. We mainly consider the merged Dirac cone graphene, which provides a suitable playground for the pseudospin assisted chirality transfer\cite{montambaux2018artificial,montambaux2009universal}.
The tight-binding approximation of the modified graphene lattice shows direction-dependent hopping that allows for a mass term that controls the transition from a topologically metallic phase into a saddle point phase where the Dirac cones merge and generate a semi-Dirac dispersion\cite{polini2013artificial,wang2015rare,montambaux2018artificial,PhysRevLett.125.186601}. The dispersion is highly anisotropic where it is linear in one direction and quadratic in the perpendicular direction, and the group velocities at the Fermi surface are asymmetric \cite{PhysRevB.80.045401,PhysRevB.80.153412,montambaux2009universal,volovik2007quantum}. Consequently, the transport properties alter dramatically, leading to anisotropic relaxation time and conductivity of the Dirac fermions \cite{adroguer2016diffusion,PhysRevB.100.035441,nualpijit2018tunable,PhysRevB.99.115406}. At energies above the gap, the Fermi surface is connected and distorted with a reach pseudospin profile that we will benefit from in our study. 

We analyse these highly anisotropic transport properties by first constructing a quantum kinetic model for the fermions obeying a general Dirac Hamiltonian with dilute disorder. We show that within the semiclassical picture, anisotropic velocities induce a rich feature on the pseudospin texture for the merged cone graphene that is otherwise trivial for the graphene. The expansion of the semi-Dirac Hamiltonian near the two Dirac cones yields two linear Dirac Hamiltonian with opposite chirality \cite{PhysRevB.80.153412,montambaux2009universal}. As we will extend in the text, this alludes to the fact that chirality is a well-defined parameter in the semi-Dirac systems at low energies. Therefore, while the pseudospin is nontrivially locked to the momentum direction, so is the valley index. Therefore, the pseudospins projected in the quadratic direction have one chirality majority and are thus valley polarized. This is crucial in attaining valleytronic applications in the semi-Dirac phase using the pseudospin index when the Dirac cones are still well defined. In a recent study, this fact is used to achieve valley polarization in the merged Dirac cone systems with the help of pseudospin tunnelling \cite{ang2017valleytronics} in the ballistic regime. 

We furthermore extend our comprehensive transport theory of the disordered modified graphene lattice for the valleytronic application using the two-band Bloch Hamiltonian, illustrating a double Dirac cone system, quadratic in one direction and linear in the other. We then construct and establish a 2D real space diffusion model for the pseudospin and the scalar charge using the kinetic model. Noting the anisotropy of the lattice, we will demonstrate the relatively weak valley mixing and hence robust valley polarization of the carriers during the charge diffusion by splitting the resultant diffusion equations into the band quadratic and linear directions.
We adopt a diffusive viewpoint where the samples are impure, and carriers are prescribed by statistical distribution functions and then study the direction-dependent dynamics of charge densities associated with each chirality using a set of drift-diffusion equations. 

The organisation of the paper is developed gradually as follows. We begin the paper by introducing a general two-band Bloch Hamiltonian describing both the Dirac and the semi-Dirac phases and constructing the transport theory using the quantum Liouville's equation in the presence of the impurity scattering in Section \ref{Sec.1}. We then solve the semiclassical Boltzmann equation by self-consistent perturbative methods to obtain a generalised distribution function in the mixed coordinates in Section \ref{Sec.3}. We drive the diffusion model by approximating the general solution in the gradient expansion, obtaining the coupled differential equations for the charge and pseudospin densities in both graphene and merged Dirac cone graphene in Section \ref{Sec.4} and conclude the paper by giving the summary in Section \ref{Sec.5}.

\section{Quantum kinetics}\label{Sec.1}
\subsection{Model}
\begin{figure}[t]
	\centering
	\includegraphics[width=.8\linewidth]{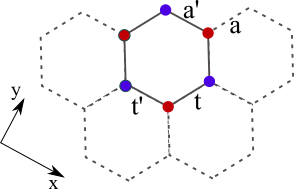}
	\caption{(Color online) Hexagonal lattice of graphene with initial isotropic sublattice distance $a=a^\prime$ and hopping parameters $t=t^\prime$ between the A (Blue) and B (Red) sublattices for the three nearest neighbours. Modifying the lattice with $a<a^\prime$ lead to weaker hopping in the $x$ and stronger hopping in the perpendicular directions. The hopping ratio deviates from one thus $\beta=t/t^\prime>1$. This, in turn, reduces the Fermi velocity (slop of the Diracness) in the parabolic direction compared to the linear spectrum in the $y$-direction.}
	\label{fig:artificial}
\end{figure}
A quantum transport equation is a quantitative description of the nonequilibrium dynamics of charge in nanostructures written in their distribution function. The distribution function reaches the Fermi-Dirac distribution when the nonequilibrium forces are turned off, the quasiparticles encounter many scattering events and finally relax into the equilibrium. We start with the Hamiltonian 
\[H=(c_x{k_x^\alpha}-\Delta)\sigma_x+c_yk_y\sigma_y,\]
describing the Hamiltonian of the standard graphene for $\alpha=1$, with linear isotropic dispersion in both directions with velocities $c_x=c_y=v_F=3ta$, whereas $\alpha=2$ pertains to the semi-Dirac system in 2D (modified graphene) having quadratic dispersion in the $x$-direction and linear on the perpendicular the $y$-direction. In this case, the parameters $c_x=3t^\prime a^2/8,~c_y=3ta$ are the inverse effective mass in the $x$-direction and effective velocity in $y$-direction, respectively, and $a$ is the lattice constant (FIG.\ref{fig:artificial}). As we will show, the parameter $\Delta>0$ is the bandgap for the case of semi-Dirac graphene with $\alpha=2$. In contrast, we set $\Delta=0$ for the standard graphene, so it is only a momentum offset and has no physical effect on our calculations. The Dirac Hamiltonian is the low energy approximation of the general semi-Dirac Hamiltonian when $\varepsilon_F\ll\Delta$. During the rest of the paper, when we discuss the case of the semi-Dirac system, we implicitly mean $\alpha=2$. In the semi-Dirac graphene, the ratio of the hopping parameters of the hexagonal lattice is bounded to $1<\beta<2$, while it is $\beta=1$ for graphene. \cite{montambaux2018artificial}

Note that by substituting
$H_0\rightarrow H/(c_y^2/c_x)$ and then replacing $\Delta\rightarrow \Delta/(c_y^2/c_x)$ and $k_{x(y)}\rightarrow k_{x(y)}/(c_y/c_x)$ we obtain the dimensionless Hamiltonian
\begin{equation}\label{eq:Ham_merge}
H_0= \boldsymbol{\sigma}\cdot\mh,
\end{equation}
where we define the vector $\mh=({k_x^\alpha}-\Delta,k_y)$ and $k_{x(y)}$ gives the components of the momentum vector, and $\boldsymbol{\sigma}=({\sigma}_x,\sigma_y)$ is the in-plane vector representing the Pauli matrices on the sublattice space. In the presence of external electric field and impurity potentials, Hamiltonian overall is represented by 
\begin{equation}\label{eq:Ham}
H_\text{total}=H_0+U(\mathbf{x})+H_\text{ext},
\end{equation}
where the scalar impurity potential is given by
\begin{equation}
U(\mathbf{x})=\lambda\sum_{\mathbf{x}_i}^{N_\text{imp}}\delta(\mathbf{x}-\mathbf{x}_i),
\end{equation}
$\mathbf{x}_i$ corresponds to the location, and $N_\text{imp}$is the number of the dilute impurities in the lattice. The impurity strength $\lambda$ is a small perturbative parameter. The impurity potential may have a matrix structure in pseudospin space, but we only consider the diagonal part and neglect the pseudospin mixing scatterings. The last term in Eq.~\ref{eq:Ham} is the interaction of the particle with the external electric field, $H_\text{ext}=\mathbf{E}\cdot\mathbf{x}$, as the source of bias driving the system out of equilibrium.
\begin{figure}[t!]
	\centering
	\includegraphics[width=1\linewidth]{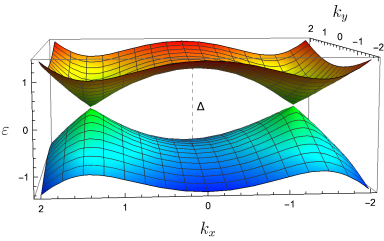}
	\caption{(Color online) Spectrum of a double Dirc cone Hamiltonian in (\ref{disp}) where the parameter $\Delta$ describes the gap term that separates the topological from a trivially gapped phase.}
	\label{fig:dispersion}
\end{figure}

The Hamiltonian (\ref{eq:Ham_merge}) gives the anisotropic dispersion as
\begin{equation}\label{disp}
\varepsilon=\pm|\mh|=\pm\sqrt{(k_x^\alpha-\Delta)^2+ k_y^2},
\end{equation}
allowing for two Dirac cones when $\Delta>0$ positioned at $\mathbf{K}=(0,\pm \sqrt{\Delta})$ (Fig.\ref{fig:dispersion}). The semi-Dirac graphene $\Delta<0$ gives a gapped dispersion; therefore, the parameter $\Delta$ also characterizes a transition from a topological phase with a double cone into a trivial phase. Right at the middle point between the Dirac cones and at $\Delta=0$ there is a saddle point where the Dirac cones merge. Throughout this study, we only consider a case with $\varepsilon_F>\Delta>0$.

Note that the velocity matrix according to the Hamiltonian in (\ref{eq:Ham_merge}) is given by 
\begin{align}
\mathbbl{v}=\nabla_\mathbf{k}H_0=(k_x^{\alpha-1}\sigma_x,\sigma_y).
\end{align}
For graphene ($\alpha=1$), we observe pseudospin-momentum locking, $\mathbbl{v}=\sigma$, where the pseudospin is parallel to the direction of the momentum vector. However, for semi-Dirac graphene where $\mathbbl{v}=(k_x\sigma_x,\sigma_y)$ the pseudospin-momentum locking is lifted, and the pseudospin has a complicated and rich texture in momentum space (FIG.\ref{fig:sudo}).

\subsection{Derivation}

To describe the collective behaviour of the carriers given by the Hamiltonian (\ref{eq:Ham}), we will use the generic definition of the density matrix where its evolution is given through the quantum Liouville's equation (setting $\hbar=1$)
\begin{equation}\label{eq:liuv}
i\dot{\rho}=[H,\rho]_-,
\end{equation}
where the distribution matrix $\rho$ is a $2\times 2$ matrix in sublattice.
Here we assume that there is no external field present at far past $t\rightarrow -\infty$, and the system is initially in equilibrium characterized by $\rho_{\text{eq}}$. Then by adiabatically turning on the external field, using the ansatz $\mathbf{E}(t)=\mathbf{E}\:e^{st}$, the system is gradually driven out of the equilibrium, however, asymptotically close to it. Within the linear response, the density matrix is expressable as a power series of the external field,
$\rho(t)=\rho_{\text{eq}}(H_0)+\delta\rho(t)$,
where the nonequilibrium part of the density matrix is linear in the electric field and requires $\delta\rho(t)=f e^{st}$. The Liouville equation in linear order reads\cite{PhysRev.108.590,*PhysRev.109.1892,culcer2008weak}
\begin{equation}\label{eq:liuv_noneq}
i\frac{\partial\delta\rho}{\partial t}=[H_0+U(\mathbf{x}),\delta\rho]_{-}+[H_{\text{ext}},\rho_0]_{-}.
\end{equation}
In the momentum basis where the Hamiltonian (\ref{eq:Ham_merge}) is diagonal and noting $\langle k|\delta\rho|k^\prime\rangle=e^{st}f_{kk^\prime}$, then the Liouville's equation reduces into a matrix equation in the momentum space
\begin{align}\label{eq:liuv_mtrx}
isf_{kk^\prime}&=H_0(k)\:f_{kk^\prime}-f_{kk^\prime}\:H_0(k^\prime)-\mathbf{E}\cdot[\mathbf{x},\rho_0]_{kk^\prime}\nonumber\\
&\qquad+\sum_{k^{\prime\prime}}\Big(U_{kk^{\prime\prime}}f_{k^{\prime\prime}k^\prime}-f_{kk^{\prime\prime}}U_{k^{\prime\prime}k^\prime}\Big).
\end{align}
The equation can be decomposed into diagonal and off-diagonal equations for $f_{kk^\prime}=f_k\delta_{kk^\prime}+f_{kk^\prime}$, such that
the diagonal part becomes
\begin{align}\label{eq:diag}
isf_k&=H_0(k)\:f_k-f_k\:H_0(k)+\sum_{k^\prime\neq k} \Big(U_{kk^\prime}f_{k^\prime k}-f_{kk^\prime}U_{k^\prime k}\Big)\nonumber\\
&\hspace{80pt}-e\mathbf{E}\cdot[\mathbf{x},\rho_0]_{kk}.
\end{align}
The off-diagonal part noting $f_k\gg f_{kk^\prime}$\cite{PhysRev.108.590,*PhysRev.109.1892} yields
\begin{align}\label{eq:off_diag1}
isf_{kk^\prime}&=H_0(k)\:f_{kk^\prime}-f_{kk^\prime}\:H_0(k^\prime)\nonumber\\
&\qquad+U_{kk^\prime}f_{k^\prime}-f_{k}U_{kk^\prime},
\end{align}
which returns the off-diagonal distribution function as
\begin{align}\label{eq;off_sol}
f_{kk^\prime}=\frac{1}{2\pi i}\int_{-\infty}^{+\infty}dz\;\mathbb{G}^{^{\scriptstyle
		R}}_k(z)\Big(U_{kk^{\prime}}f_{k^\prime}-f_{k}
U_{kk^{\prime}}\Big)\mathbb{G}^{^{\scriptstyle
		A}}_{k^\prime}(z),
\end{align}
where we introduced retarded and advanced Green's functions 
\begin{align}\label{eq:resolvent}
\mathbb{G}^{^{\scriptstyle
		R/A}}_k(z) = \Big(z \pm is/2-H_0(k)\Big)^{-1}.
\end{align}

Next, we substitute this solution back in the diagonal equation (\ref{eq:diag}) to obtain the transport equation \cite{PhysRev.108.590,*PhysRev.109.1892,culcer2008weak,culcer2017interband}
\begin{align}\label{eq:Q_transport}
s f_k
+i[H_0(k)\:,f_{k}]-e\mathbf{E}\cdot[\nabla_k,f]_k=\mathbb{I}[f_k].
\end{align}
The reduced Quantum Liouville's equation (\ref{eq:Q_transport}) shares similarities with the classical Boltzmann equation except for the second term on the left side, which involves a commutator as the quantum correction to the classical Boltzmann equation. This term describes the quantum interference effects. 
The first and the third term give the time evolution and the drift of the carriers due to the electric field, respectively. 

\subsection{Wigner function and the collision integral}

The right side of Eq.~(\ref{eq:Q_transport}), on the other hand, describes the collision integral due to the uncorrelated random impurity potential given by
\begin{align}
\mathbb{I}[f_k]&=\frac{1}{2\pi}\int
dz\;\sum_{\textbf{k}^{\prime}}
|{U}_{\textbf{k}\textbf{k}^{\prime}}|^2\times\nonumber\\
&\bigg(\mathbb{G}^{{\scriptstyle R}}_{\textbf{k}^{\prime}z} (f_{\textbf{k}^{\prime}}-f_{\textbf{k}})\mathbb{G}^{{\scriptstyle
 		A}}_{\textbf{k}z}+\mathbb{G}^{{\scriptstyle	R}}_{\textbf{k}z}
 	(f_{\textbf{k}^{\prime}}-f_{\textbf{k}})\mathbb{G}^{{\scriptstyle
 		A}}_{\textbf{k}^{\prime}z}\bigg),
\end{align}
resembling the generalized Fermi's golden rule.
In principle, the semiclassical interpretations depict the carriers as propagating wave packets having definite momentum and moving inside the crystalline lattice in the real space. This requires constructing a quantum mechanical phase space using the Wigner transform in which both the position and momentum  of the carriers will be addressed. Based on this observation, we now apply the Wigner transform to the Eq.~(\ref{eq:Q_transport}) followed by a Laplace transform to restore the time dependency. After simplifying the collision integral, this leads to the following semiclassical transport equation
\begin{align}\label{eq:semi}
	&\partial_tf-\mathbf{E}\cdot\nabla_\mathbf{k}f+i[H_0,f]_{_{-}}+\frac{1}{2}[\mathbbl{v}\:,\:\nabla_\mathbf{x}f]_{_+}\nonumber\\
	&\hspace{.3in}=i\frac{\rho}{\tau}-\frac{f}{\tau},
\end{align}
where the identification $f = f(\mathbf{k},\mathbf{x},t)$ implies the Wigner (semiclassical) distribution function and
\begin{equation}\label{eq:realDist}	
\rho_{\varepsilon}(\mathbf{x})=\frac{1}{2\pi i}\sum_{\textbf{k}}
\left(\mathbb{G}^{{\scriptstyle
		R}}_{\textbf{k},\varepsilon}f_{\textbf{k}}-f_{\textbf{k}}
\mathbb{G}^{{\scriptstyle
		A}}_{\textbf{k},\varepsilon}\right).
\end{equation}
Compared to the quantum transport equation (\ref{eq:Q_transport}), the only difference in the semiclassical equation (\ref{eq:semi}), other than the collision terms, is the fourth term as the diffusion term originating from the Poisson bracket. The collision terms are reduced using the relaxation time approximation where $\tau^{-1}=\sum_{\textbf{k}}|{U}_{\textbf{k}\textbf{k}^{\prime}}|^2\delta(\varepsilon_k-\varepsilon_F)$.

The Boltzmann equation is usually solved perturbatively in the electric field to obtain the transport coefficients. In the next section, however, to study the interplay of the charge and the pseudospin, we will further reduce the semiclassical equation into the drift-diffusion model by gradient expansion of the position-dependent densities.

\section{Generalized kinetic distribution}\label{Sec.3}

This simplification of the transport equation will be carried out by introducing a general distribution matrix as
\begin{equation}
\mathfrak{g}_{\textbf{k},\varepsilon}=\frac{1}{2\pi i}\Big(\mathbb{G}^{{\scriptstyle
		R}}_{\textbf{k},\varepsilon}f_{\textbf{k}}-f_{\textbf{k}}
\mathbb{G}^{{\scriptstyle
		A}}_{\textbf{k},\varepsilon}\Big),
\end{equation}	
which gives the distribution of the carriers at fixed energy, hence the solution of the stationary and charge uniform transport equation \cite{PhysRevLett.93.226602,PhysRevLett.95.256602}. This particular form of $\mathfrak{g}_{\textbf{k},\varepsilon}$ is directly related to the Keldysh (kinetic) Green's function, which defines the mass shell distribution function  \cite{kamenev2011field,*rammer2007quantum}. The real space charge distribution at energy $\varepsilon$ is then given by summing over the momentum variables
\begin{equation}\label{eq:realDist1}	
\rho_{\varepsilon}(\mathbf{x})=\sum_{\textbf{k}}
\mathfrak{g}_{\textbf{k},\varepsilon},
\end{equation}
where we parameterize $2\times 2$ real space distribution matrix $\rho(\mathbf{x})=n+\boldsymbol{\sigma}\cdot\mathbf{s}+{\sigma_zs_z}$, where $n(\mathbf{x})$ is the nonequilibrium scalar, $\mathbf{s}=(s_x(\mathbf{x}),s_y(\mathbf{x}))$ and $s_z(\mathbf{x})$ are the nonequilibrium pseudospin charge distributions. We tend to reduce the Boltzmann equation into a balanced equation for the local distribution matrix $\rho(\mathbf{x})$.

Multiplying the semiclassical transport equation (\ref{eq:semi}) from the left and right sides by the appropriate Greens function, followed by subtracting them, gives
\begin{equation}\label{eq:genral_kinetic}
(s+\tau^{-1})\mathfrak{g}_{\textbf{k},\varepsilon}+i[H_0,\mathfrak{g}_{\textbf{k},\varepsilon}]_-=\mathfrak{L}_{\textbf{k},\varepsilon}.
\end{equation}
The left side can be understood as the fast relaxation into the equilibrium which is mediated with the terms on the right hand side as scattering and anisotropic deviation from the local distribution given by  $\mathfrak{L}_{\textbf{k},\varepsilon}=\mathfrak{L}_{\textbf{k},\varepsilon}^{(0)}+\mathfrak{L}_{\textbf{k},\varepsilon}^{(1)}$ where
\begin{align}\label{eq:sors_1}
\mathfrak{L}_{\textbf{k},\varepsilon}^{(0)}&=i\tau_{k}^{-1}\big(\mathbb{G}^{{\scriptstyle\:R}}_{\textbf{k}\varepsilon}\rho_\varepsilon-\rho_\varepsilon\mathbb{G}^{{\scriptstyle\:A}}_{\textbf{k}\varepsilon}\big),\\
\label{eq:sors_2}
\mathfrak{L}_{\textbf{k},\varepsilon}^{(1)}&=-\frac{1}{2}\big[\mathbbl{v};\nabla_x\,\mathfrak{g}_{\textbf{k},\varepsilon}\big]_+.
\end{align}
The solution to the transport equation~(\ref{eq:genral_kinetic}) can be written as
\begin{equation}
\mathfrak{g}_{\textbf{k},\varepsilon}=\oint\:\frac{dz^\prime}{2\pi}\;\textnormal{G}_{\textbf{k}z^\prime}^+\;\mathfrak{L}_{\textbf{k},\varepsilon}\;\textnormal{G}_{\textbf{k}z^\prime}^-,
\end{equation}
where the Greens functions are the same as defined in (\ref{eq:resolvent}) with small modification  $s\rightarrow\Omega=s+1/\tau$. The integrand have 4 simple singularities, two denoted by $z^\prime=\pm\varepsilon_k-i\Omega/2$ lie in the retarded (lower) plane and the other two denoted by $z^\prime=\pm\varepsilon_k+i{\Omega}/{2}$ reside in the advanced (upper) plane. Computing the $z^\prime$ integral over either of the half planes consequently gives
\begin{align}\label{eq:general_sol}
\mathfrak{g}_{\textbf{k},\varepsilon}
&=\zeta_1\;\mathfrak{L}_{\textbf{k},\varepsilon}+i\zeta_2\;[\boldsymbol{\sigma}\cdot{{\hat{\mh}}},\mathfrak{L}_{\textbf{k},\varepsilon}]_{_-}
+\zeta_3\;\boldsymbol{\sigma}\cdot{{\hat{\mh}}}\;\mathfrak{L}_{\textbf{k},\varepsilon}\;\boldsymbol{\sigma}\cdot{{\hat{\mh}}}\nonumber\\
&=\mathbb{L}[\mathfrak{L}_{\textbf{k},\varepsilon}],
\end{align}
where $\hat{\mh}=\mh/\varepsilon$. We specify the coefficients as follows
\begin{equation}\label{coefficients}
\zeta_1=\frac{\tau(1+2\gamma^2)}{1+4\gamma^2},\quad
\zeta_2=\frac{2\gamma\tau}{1+4\gamma^2},\quad
\zeta_3=\frac{2\gamma^2\tau}{1+4\gamma^2},
\end{equation}
and $\gamma=\varepsilon_{F}\tau$ and in the quasistationary regime $\Omega\approx1/\tau$. We can then establish the different corrections by iteration to get
\begin{align}\label{perturb}
\mathfrak{g}_{\textbf{k}\varepsilon}^{(0)}&=\mathbb{L}[\mathfrak{L}_{\textbf{k},\varepsilon}^{(0)}(\rho)],\\
\label{perturb2}
\mathfrak{g}_{\textbf{k}\varepsilon}^{(i)}&=\mathbb{L}[\mathfrak{L}_{\textbf{k},\varepsilon}^{(1)}(\mathfrak{g}_{\textbf{k}\varepsilon}^{(i-1)})] \qquad i\geq1.
\end{align}
\begin{figure}[h!]
	\centering
	\includegraphics[width=.9\linewidth]{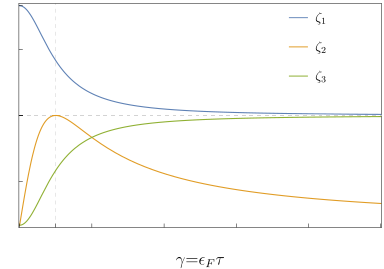}
	\caption{(Color online) The values for the different coefficients in (\ref{coefficients}) with respect to the $\gamma=\varepsilon_F\tau$.}
	\label{fig:iso}
\end{figure}
The second term in the first line of Eq. (\ref{eq:general_sol}) represents the pseudospin precession about the effective magnetic field $\hat{\mh}$. In the semi-Dirac phase, due to the anisotropy of the dispersion relation, the pseudospin-momentum locking is lifted, and in the presence of interference effects, pseudospin can have an arbitrary direction in momentum space. It is generally possible to decompose the pseudospin distribution into the parallel and perpendicular components with respect to the vector $\mh$, within the $\mh-\mathbf{s}$ plane, i.e, $\mathbf{s}=(\mathbf{s}\cdot\mh)\mh+\mh\times(\mh\times\mathbf{s})$. The first term on the right side of the equality is parallel to $\mh$ thus commutes with the Hamiltonian and therefore is a conserved quantity. On the other, the perpendicular component is non-conserved and undergoes precession which is proportional to the $z$-component of the pseudospin $[\mh\cdot\sigma,\mathbf{s}_\perp\cdot\sigma]_-=\mh\times\mathbf{s}\:\sigma_z$. Note that the $z$-copmponet of the pseudospin characterizes the coherence and the electron propagation between the two sublattices\cite{li2020artificial,PhysRevB.87.165131,culcer2008weak,trushin2011pseudospin}. This result indicates that the non-conserved pseudospin components are an intrinsic and scattering-independent feature in systems possessing pseudospin and affects the charge conduction and diffusion\cite{culcer2008weak}. Furthermore, in the clean regime where $\gamma\gg 1$ the coefficient of the precessional term, $\zeta_2\propto\varepsilon_F^{-1}$, is independent of the relaxation time $\tau$ so scattering-independent, while the others are linear in $\tau$, i.e., $\zeta_1,\zeta_3\propto\tau$ (Fig. \ref{fig:iso}). 

We are interested in the quasi-stationary system where $s\tau\ll1$, hence, i.e., $\Omega\approx\tau^{-1}$, i.e., we neglect $s$ (the Fourier conjugate of the time) in the quasi long-time stationary transport regime. This limit indicates that the scattering rates due to the impurity are so short, $\tau^{-1}\propto\lambda^2$, that it takes a long time for the system to reach equilibrium, indicating a dilute impure system. We adopt this limit where $\zeta_1=\zeta_3\approx1/2,~ \zeta_2\approx 0$ to study the coupled dynamics of charge and pseudospin. 

We explore the charge and pseudospin transport in both graphene and merged cone graphene and their role in valley transport by deriving the coupled charge-pseudospin diffusion equations \cite{PhysRevLett.93.226602,PhysRevLett.95.256602}.

\section{Gradient expansion and Diffusion equation}\label{Sec.4}

 The general form of the distribution function using (\ref{eq:general_sol}) can be written as
\begin{equation}
\mathfrak{g}_{\textbf{k}\varepsilon}^{(i)}={\Lambda}^{(i)}(\hat{\mh})+\boldsymbol{\Gamma}^{(i)}(\hat{\mh})\cdot\boldsymbol{\sigma}+\Gamma_z^{(i)}\sigma_z,
\end{equation}
where $i$ indicates the iteration order. 
The full description of the coefficients $\Lambda^{(i)}$ and $\boldsymbol{\Gamma}^{(i)}$ for each iteration and their functional form is provided in Appendix A. The zeroth correction will be proportional to $2\times 2$ distribution matrix $\rho(\mathbf{x})$, while the first and second corrections produce their first and second gradients. The general solution (\ref{eq:general_sol}) can then be represented as a sum of the gradient expansion of the distribution matrix $\rho$, such that
\begin{align}
\mathfrak{g}_{\textbf{k}\varepsilon}=\mathfrak{g}_{\textbf{k},\varepsilon}^{(0)}+
\mathfrak{g}_{\textbf{k},\varepsilon}^{(1)}+
\mathfrak{g}_{\textbf{k},\varepsilon}^{(2)}=\mathcal{F}_{\textbf{k}\varepsilon}(\rho,\nabla\rho,\nabla^2\rho).
\end{align}
At this stage, according to Eq.~(\ref{eq:realDist}), we integrate out the momentum degrees of freedom, giving a matrix identity 
\begin{equation}\label{iden}
\big\langle\mathcal{F}_{\textbf{k}\varepsilon}(\rho,\nabla\rho,\nabla^2\rho)\big\rangle_{\textbf{k}} \equiv\rho,
\end{equation}
where the bracket notation stands for the two dimensional momentum integration. We make use of the transformation $(h_x,h_y)\rightarrow (\varepsilon,\theta)$ such that
\begin{equation}
k_x^\alpha+\Delta = \varepsilon\cos\theta,\hspace{.3in}k_y = \varepsilon\sin\theta,
\end{equation}
which transforms the integrals into
\begin{equation}\label{jacobian}
\langle{\cdots}\rangle_\textbf{k}
=\int_0^{\varepsilon_F} d\varepsilon\; \delta(\varepsilon-\varepsilon_F)\int_{-\theta_0}^{\theta_0}\; \frac{d\theta}{2\pi}\:||J||\:\cdots.
\end{equation}
The Jacobian of the transformation for graphene reads $||J||= 2\varepsilon$ with $-\pi<\theta_0<\pi$, and for the semi-Dirac graphene is $||J||=\sqrt{\varepsilon/2(\cos\theta-\eta)}$ with $-\cos^{-1}\eta<\theta_0<\cos^{-1}\eta$, and $\eta=\Delta/\varepsilon$.
%\[ ||J|||=\frac{(2\varepsilon)^{1/\alpha}}{\alpha} (\cos\theta-\eta)^{\frac{1-\alpha}{\alpha}}\]
We consider the connected Fermi surface for the merged Dirac cone where $\varepsilon_F\gg\Delta$ and $\eta\rightarrow0$.

\subsection{charge-pseudospin diffusion in graphene}

In graphene where $\alpha=1$, computing the perturbative solutions and integrating over the momenta (Appendix A) in the limit $\gamma\gg1$, the identity (\ref{iden}) returns
\begin{align}
D\:\nabla^2n-v_F\:\nabla\cdot\mathbf{s}&=0,\;\label{ch:diff}\\
\frac{1}{2}D\:\nabla^2\mathbf{s}+ D\:\nabla(\nabla\cdot\mathbf{s})-{v_F}\:\nabla n&=\frac{\mathbf{s}}{\tau}.\label{s12:diff}
\end{align}
These equations show a strong charge-pseudospin coupling and, at the same time, they bear similarities with the diffusion equations that have been introduced for the spin-charge coupled dynamics in the disordered two-dimensional electronic systems with spin-orbit coupling\cite{PhysRevLett.95.256602, PhysRevLett.93.226602} and the surface of 3D topological insulators\cite{PhysRevLett.105.066802}. These similarities indicate that a portion of the charge behaves quite differently from the scalar charge; thus, it can be polarized, similar to the spin. Furthermore, in a stationary system, the continuity equation combined with the equation (\ref{ch:diff}) enforces the relation
\begin{equation}\label{eq:conti}
\mathbf{J}_\text{ch}=D\:\nabla n-v_F\:\mathbf{s},
\end{equation}
and note that, in the presence of both the electric and chemical potential bias, we shall adopt the boundary condition $\nabla\rightarrow\nabla-e\:E\partial_\varepsilon$.
This quick result indicates that, in addition to the electrochemical potential contribution to the current, which induces the drift-diffusion of the charge density, there is yet another contribution stemming from the in-plane polarization of the pseudospin charge or the admixture of the charge from both the sublattices. The contributions are comparable by looking at their coupling constants, as the gradient concentration couples with $D=v_F\ell/2$, the pseudospin contributes with constant $v_F$, and $\ell$ is the mean-free path of the particles. In 2D electronic systems with spin-orbit interaction, a similar relationship exists between the charge density and the in-plane polarization of the spin density transverse in the transport direction \cite{PhysRevLett.104.116401,PhysRevLett.105.066802} which manifests the bias-induced excess spin chemical potential or the spin accumulation in the system. Therefore (\ref{eq:conti}), in particular, means that the majority of the pseudospin charge have polarization parallel (in contrast to the spin version) to the direction of the transport. Note that the in-plane pseudospin states are the superposition of the electronic states in the two sublattices $|\sigma_{x(y)}\rangle=(|A\rangle+e^{i\phi_{x(y)}}|B\rangle)/\sqrt{2}$, where the sublattices are the eigenstates of the out-of-plane pseudospin matrix, i.e., $|A(B)\rangle=|\sigma_z,+(-)\rangle$, and $\phi_x(\phi_y)=\pi(\frac{\pi}{2})$ is the azimuthal angle in the equator on the pseudospin Bloch sphere.
 It is fruitful to rewrite (\ref{s12:diff}) in the form of the continuity equation as $\partial_i{J}_{i,s_j}-{v_F}\partial_i n={s}_i/\tau$, where now ${J}_{i,s_j}=\frac{1}{2}D\partial_i{s}_j+D\partial_j{s}_i$ is the pseudospin current. Note that, similar to the spin, the continuity equation for the pseudospin is not conserved. As we discussed earlier, the presence of non-conserved pseudospin is intrinsic in graphene and it contributes to the peculiar electronic properties of Dirac Fermions due to the quantum coherence present in the Hamiltonian \cite{culcer2008weak,culcer2017interband}.
\begin{figure}[t!]
	\centering
	\includegraphics[width=.8\linewidth]{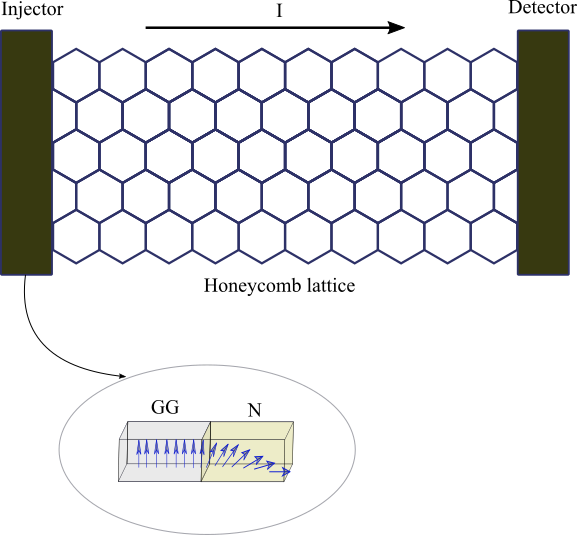}
	\caption{(Color online) Experimental setup for observation of the pseudospin assisted valley transport.  While the current $I$ flow into the system in the $x$-direction, the in-plane pseudospin is injected through the left interface into the scattering region with a honeycomb lattice structure. The longitudinal pseudospin density $s_x$ gradually increases according to (\ref{sx:density}) and eventually piles up at the right interface. The electric signal of this accumulation can be seen as a voltage drop between the right and the left interfaces}
	\label{fig:circuit}
\end{figure}

Solving a general two-dimensional diffusion model such as in (\ref{ch:diff}) and (\ref{s12:diff}) is a cumbersome task and requires complete knowledge of the boundary conditions. We are only interested in designing a setup in quasi-one dimension to study the diffusion of the pseudospin densities in transport direction\cite{majidi2011pseudospin}. To utilize the diffusion equations, we now assume that the graphene nanoconductor is set along the $x$ direction and a constant current $I$ flowing through it (FIG.\ref{fig:circuit}), then
\begin{align}\label{dd_1}
J_\text{ch}=D\partial_xn-v_Fs_x=\frac{I}{e},\\
\frac{3}{2}D\partial_x^2s_x-v_F\partial_xn-\frac{s_x}{\tau}=0.
\end{align}
We furthermore assume that we manage to withhold a degree of in-plane pseudospin polarization $\kappa$ generated in the left interface, which decays inside the conductor and vanishes at the right boundary. This can be done by implementing a junction made of a gapped monolayer of graphene(GG), with $H_{GG}=v_F\:\sigma\cdot\mathbf{k}+\delta \sigma_z$ where $\delta$ is the substrate potential inducing band gap, in proximity to normal graphene monolayer (N). The eigenstates for the gapped graphene region then read
\begin{align}\label{gg_eigen}
|\psi_{\varepsilon_{GG}}\rangle=e^{ik_x x}
\begin{pmatrix}
e^{i\phi/2}\\e^{-\beta}\:e^{\pm i\phi/2}
\end{pmatrix}
\end{align}
where $\phi=\tan^{-1}(\frac{k_y}{k_x})$ and $\beta=\tanh^{-1}(\delta/\varepsilon_{GG})$. Note that when $\delta=0$ then $\beta=0$ and we recover the usual solution for the normal graphene in (\ref{gg_eigen}). Using the eigensolutions (\ref{gg_eigen}) the pseudospin polarization can be computed as
\begin{align}
\langle\boldsymbol{\sigma}\rangle_{\psi_{\varepsilon_{GG}}}=\sqrt{1-\left(\frac{\delta}{\varepsilon_{GG}}\right)^2}\:\mathbf{k}_{||}+\frac{\delta}{\varepsilon_{GG}}\:\mathbf{k}_\perp,
\end{align}
where $\mathbf{k}_{||}$ is the momentum of Dirac electrons in GG and $\mathbf{k}_{\perp}$ is the out of the plane direction. Note that, by tuning $\delta\approx\varepsilon_{GG}$, then the pseudospin is completely out of the plane and in the $z$-direction in the GG region inside the injector. At the interface between GG/N, when the substrate potential drops $\delta\rightarrow 0$, due to the proximity effects, the pseudospin rotates into the in-plane along the propagation direction.

In earlier works, a similar setup was proposed using gapped graphene junctions and phosphorene ribbons to induce pseudospin polarization at the interface\cite{majidi2011pseudospin,soleimanikahnoj2017pseudospin,majidi2013quantum}. Therefore for the pseudospin current, we can introduce the following boundary conditions
\begin{align}\label{bc}
J_{s_x}\Big|_{x=0}=-\kappa\:J_\text{ch},\qquad J_{s_x}\Big|_{x=L}=0.
\end{align}
\begin{figure}[b!]
	\centering
	\includegraphics[width=1\linewidth]{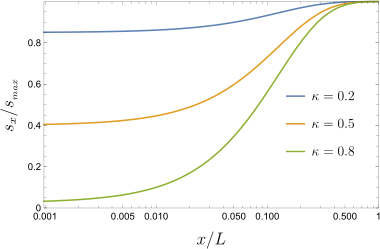}
	\caption{(Color online) Diffusion of the pseudospin density $s_x$ inside the graphene sample and at the boundary for different polarization degrees $\kappa$. We set $s_\text{max}=s_x(x=L)$.  We used $v_F=2.5\times 10^6 m/s$, $\ell=1\:nm$ for a long conductor $L=20\: nm$.}
	\label{fig:sx-accum}
\end{figure}
Then the diffusion of the pseudospin density $s_x$ is governed by
\begin{equation}\label{eq:psudo_diff}
\partial_x^2s_x-\frac{8}{3\ell^2}s_x-\frac{8I}{3ev_F\ell^2}=0,
\end{equation}
which in light of the boundary conditions (\ref{bc}) solves as
\begin{align}\label{sx:density}
s_x(x)=\frac{I}{3ev_F} \left(2-3\kappa\frac{\cosh \left(\frac{2(L-x)}{\ell }\right)}{\sinh\left(\frac{2L}{\ell }\right)}\right).
\end{align}
We observe that pseudospin density has two contributions. The second term is the injected pseudospin charge from the left electrode, which enhances along the conductor. In contrast, the first term, is the pseudospin density, $s_x(x)\propto I=\text{constant}$, inside the system induced by the electric field along the $x$ axis (Fig.\ref{fig:sx-accum}). This term is constant and nonequilibrium and independent of the disorder thus pertains to the non-conserved pseudospin density. Moreover, as a result of competition between the first and the second terms in (\ref{sx:density}), gradual enhancement of the pseudospin accumulation is observed maximising at the right interface. The elevation is more pronounced when the injected charge at the left interface is fully pseudospin polarized rather than partially polarized (Fig.\ref{fig:sx-accum}).
The difference in the pseudospin accumulation among the two interfaces thus induces a voltage drop between them. This effect is well known in spin transport, where the proximity effects in magnetic nanojunctions lead to pure spin accumulation with long precession length and increased spin detection sensitivity\cite{fukuma2011giant,PhysRevLett.102.036601}. Equation (\ref{sx:density}), therefore, indicates an all-electric generation of the pure pseudospin population and constitutes the main result of this paper. Similar robust in-plane pseudospin density was also reported in Ref.\cite{majidi2011pseudospin} whereupon injecting out-of-plane pseudospin polarization a non-decaying in-plane polarization is generated parallel to the direction of the electric bias at the interfaces, namely $\langle\sigma_x\rangle$, while other polarization directions inside the sample oscillate and decay on the few order of the Fermi length, $\langle\sigma_y\rangle,\langle\sigma_z\rangle\rightarrow0$. This clearly manifests the reflectionless Klein tunnelling ubiquitous in the Dirac system\cite{ni2018spin,majidi2011pseudospin,allain2011klein}. These studies utilize the eigenstate formalism to investigate the pseudospin dynamics by commuting the transmission probabilities, whereas we use the drift-diffusion model to provide the real-space distribution of the pseudospin components. The connection can, nevertheless, be elucidated by noting that $\langle\sigma_x\rangle=\text{tr}(\sigma_x\:\rho)\propto s_x$: the pseudospin polarization in eigenstate formalism is only the reiteration of the pseudospin density at that particular direction within our notation.

Our result demonstrates that the current in graphene is pseudospin in-plane polarized parallel to the direction of the transport and accumulates at the boundary of the sample. This suggests a new method for detecting the pseudospin induced electric signatures by measuring the voltage drop between the two interfaces. Now substituting the solution (\ref{sx:density}) into (\ref{dd_1}) helps to compute the voltage drop as
\begin{align}
V=-\frac{1}{e \nu}\int_0^L\:dx\:\left(\frac{\partial n}{\partial x}\right)=\frac{I}{e^2}\left({\frac{2L}{3 k_F\ell}}+ {\frac{\kappa}{k_F}}\right),
\end{align}
where $\nu=\varepsilon_F/2\pi v_F^2$ is the density of states. This result is similar to the one reported for the topological insulators\cite{PhysRevLett.105.066802} and delineates the electronic signature of the pseudospin polarization. The first term, evidently, is the ohmic contributions to the resistance proportional to $L$, the length of the conductor. Note that the term $k_F\ell\approx v_F^2/(n_\text{imp}\lambda^2)$ is independent of the Fermi energy for the short-range impurity model, as in our study, and $\lambda$ is the impurity strength\cite{PhysRevLett.98.076602}. The second term is, however, new contributes to the voltage due to the charge-pseudospin coupling (\ref{eq:conti}) and depends only on the polarization degree $\kappa$ of the injected current and the Fermi surface property $ k_F$. Therefore this term can be enhanced via external means, such as a gate voltage, to tune the Fermi energy, while the first term will not respond to this tuning and stays constant. Besides, the polarization-dependent term is independent of the choice of the boundary condition; that is, instead of (\ref{bc}), a general condition for the pseudospin density and its gradients will only alter the numerical coefficients.

\subsection{Semi-Dirac dispersion}
As discussed in Section (2), the transport is highly anisotropic for the modified honeycomb lattice with semi-Dirac dispersion. The valley degeneracy is broken in the quadratic direction, and the pseudospin densities carry net valley information. Therefore, in this case, a net pseudospin accumulation pertains to valley polarized carriers. We implement our drift-diffusion model to inspect the real space dependence of the pseudospin density inside a conductor, along the $x$-direction, with a semi-Dirac dispersion. We subtract the coupled diffusion equations (details in Appendix.\ref{App.B}) in the quadratic direction for $s_x$. Neglecting terms of the lower order, we find a simple result showing a relationship between the charge current and the charge and pseudospin relaxations (see e.g., Eq. \ref{eq:B8})
\begin{align}\label{eq:conti_semi}
v_{F,x}\:\partial_xn\approx-\frac{n-s_x}{\tau}.
\end{align}
This result shows that a balance equation holds between the diffusion of the charge and relaxation of the pseudospin in the semi-Dirac dispersion. 

Interestingly, one finds that this result is in accord with Eq. (\ref{eq:conti}) by noting that the current density is the charge deviation in units of the characteristic time $J_\text{ch}=\frac{\delta n}{\tau}$. Therefore, in (\ref{eq:conti_semi}) on the right-hand side, the term $\frac{n}{\tau}$ is the charge density deviation at the unit of relaxation time; thus, it is proportional to a drift current caused by voltage bias. These lines of arguments confirm that (\ref{eq:conti_semi}) indicates a charge-pseudospin coupling in a modified honeycomb lattice given by the semi-Dirac Hamiltonian, similar to the Dirac systems. (\ref{eq:conti_semi}) thus suggesting that such couplings are an intrinsic feature of systems with sublattice structure. 

Next, suppose the distance that takes charge to deviate from the local equilibrium is in the order of the mean free path $\ell$ such that $\partial_xn\approx\delta n/\ell$. In that case, using (\ref{eq:conti_semi}) and noting $v_{F,x}=\ell/\tau$, we immediately find $n+\delta n\approx s_x$, indicating that the nonequilibrium charge density is proportional to the pseudospin, showing another manifestation of the charge-pseudospin coupling. The outcome is intuitive as the pseudospin polarization (in-plane), in a real sense, is the redistribution of the electronic densities between the two sublattices. 
This, in addition, helps to recast (\ref{eq:conti_semi}) as a differential equation for $n$ as
$v_{F,x}\:\partial_xn=-\frac{\delta n}{\tau}$, which returns the solution inside the scattering region as $n(x)\sim e^{-{x}/{\ell}}$. 

At this stage, using the ansatz solution for $n(x)$, we reduce the coupled system of equations for the semi-Dirac system in Appendix.B, and obtain the differential equation for $s(x)$ as
\begin{align}\label{eq:pde}
\partial_x^2s_x-\frac{1.12}{\ell}\partial_xs_x+\frac{1.52}{\ell^2}s_x=\frac{0.64}{\ell^2}\: e^{-\frac{x}{\ell}}.
\end{align}
\begin{figure}[t]
	\centering
	\includegraphics[width=1\linewidth]{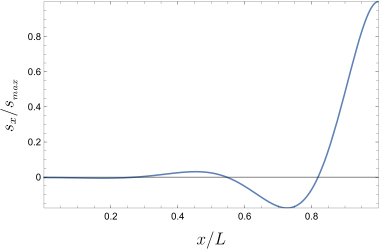}
	\caption{(Color online) Diffusion of the pseudospin density $s_x$ inside the modified honeycomb lattice given by the semi-Dirac Hamiltonian. We used $v_F= 10^6 cm/s$, $\ell=1\:nm$ for a long conductor $L=20\:nm$.}
	\label{fig:smdirac-accum}
\end{figure}
 We solve this differential equation using the boundary condition (\ref{bc}), where the constant charge current flows through the sample in the $x$-direction and pseudospin current is sustained by the injector consisting of GG/N junction at the left boundary. Since the analytical formulae for the semi-Dirac case are lengthy and too complicated, we, therefore, use numerical techniques\cite{Mathematica,*gradshteyn2014table} to solve (\ref{eq:pde}) and summarize our results in Fig.(\ref{fig:smdirac-accum}).  
 
\begin{figure}[t]
	\centering
	\includegraphics[width=.9\linewidth]{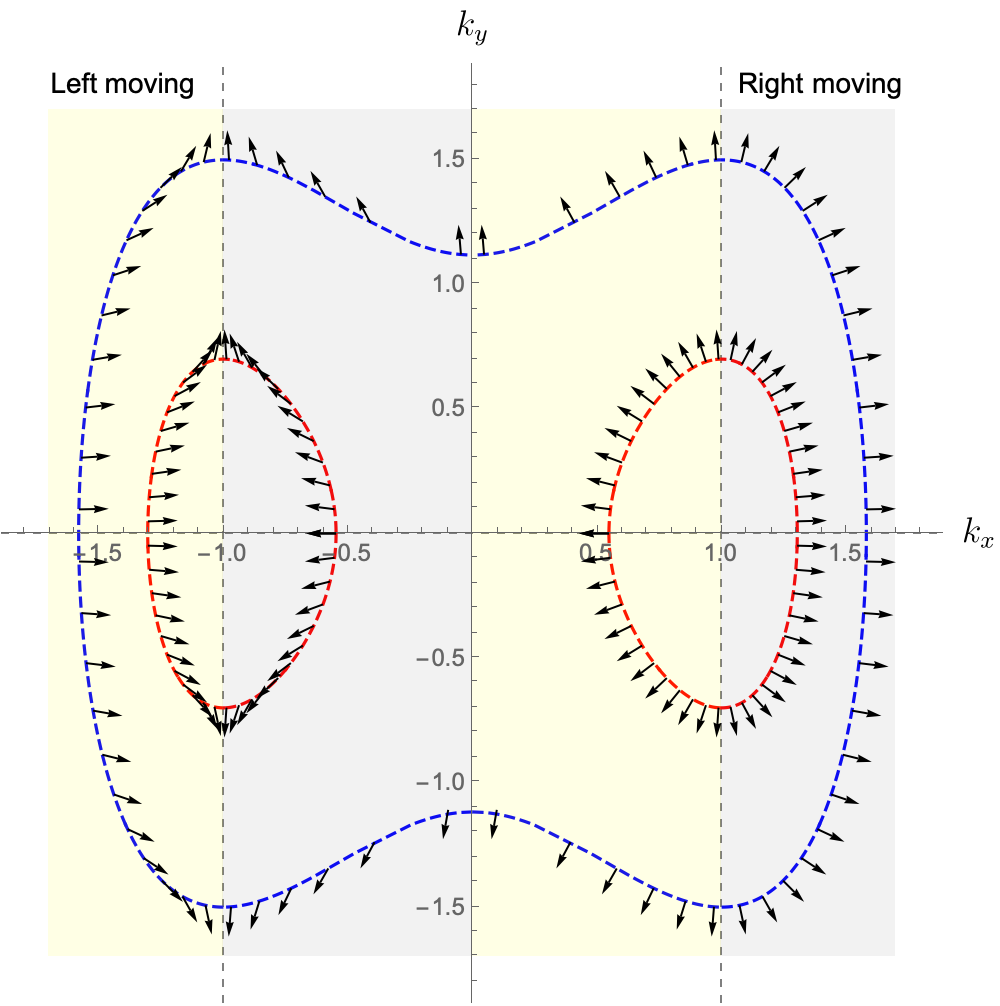}
	\caption{(Color online) Pseudospin texture of the engineered graphene with the semi-Dirac dispersion along the linear ($k_y$) and quadratic ($k_x$) directions. The Blue (Red) dashed curves show the semi-Dirac case for the 		merging (isolated) Dirac cones. The Grey (Yellow) shaded region indicates the portion of the Fermi surface with the right (left) propagating modes with anisotropic pseudospin vectors. }
	\label{fig:sudo}
\end{figure}

 Our results show that the pseudospin current is generated at the left boundary, where the accumulation is minimum. This can be justified as the reflection of the reservoir-system boundary condition where a large macroscopic reservoir imposes rapid equilibration of a nonequilibrium population (either charge, spin, or pseudospin) near the boundary\cite{blaas2021asymmetric}. The pseudospin density, similar to the case of graphene, gradually increases with the distance from the left interface and consequently saturates at the right interface (Fig.\ref{fig:smdirac-accum}). 
 
 Most remarkably, we find that the enhancement is, although qualitatively similar to graphene, however, insensitive to the degree of the injected polarization $\kappa$. This can be understood by referring to FIG.\ref{fig:sudo} where the nontrivial pseudospin texture shows intrinsic anisotropy in the semi-Dirac direction. Namely, in the merging cone limit (dashed blue curve), at the Fermi surface, the right moving propagations (grey shaded region in FIG.\ref{fig:sudo}) possess a net majority of pseudospin pointing in the $x$-direction and thus carry an intrinsic polarization. This principally indicates that materials with semi-Dirac dispersion are ideal for pseudospintronics and can operate self-sufficiently, independently of the external injection.
 
This renders an essential functionality of the semi-Dirac phase for the pseudospin-assisted valley transport by noting that in the quadratic direction, the pseudospin is nontrivially coupled to the valley due to the shape of the Fermi surface. The pseudospin profile of the Fermi surface generates pseudospin accumulation at the boundary, which can then be extracted and principally used as the source for valley polarization and valleytronic applications.

\section{Conclusion}\label{Sec.5}

We studied the charge and pseudospin coupled dynamics in graphene and its semi-Dirac version, wherein both massless and massive dispersions occur in the perpendicular momentum directions. We showed that the Fermi surface in the massive case possesses a nontrivially rich texture where the right and left propagating modes, having opposite group velocities, possess net pseudospin polarization and, in contrast to the standard graphene, at the same time carry valley information which designates them especially suitable for valleytronic applications. We demonstrate that this behaviour is mainly due to the cancellation of the pseudospin components along the transport directions on the connected Fermi surface in the semi-Dirac case. While, in the massless direction, the pseudospin components cancel each other, in the massive perpendicular direction, this cancellation is non-zero, thus resulting in a net degree of pseudospin polarization stemming from a specific valley. Furthermore, for transport in the massive direction, the portion of the anisotropic connected-Fermi-surface which contributes to the transport is asymmetric around the two Dirac points, thus giving rise to the net valley population of the propagating modes. 

We construct a quantum kinetic model for the density matrix in leading order in the impurity potential to corroborate the coherent pseudospin and valley dynamics. By integrating the momentum degrees of freedom, we finally obtain the real space drift-diffusion equations describing the coupled dynamics between the charge and the pseudospin. The particle current consists of a pseudospin part that induces a novel magnetoelectric effect in graphene and semi-Dirac graphene, other than the electrochemical potential gradient contribution. By solving the 1D diffusion equation and obtaining the form of the pseudospin density profile, we show that the pseudospin gradually enhances and piles up at the end boundary, thus signifying a voltage drop between the two interfaces as a definite consequence of the charge-pseudospin coupling. In graphene, however, the accumulated pseudospins do not have a net valley character due to the isotropy of the Fermi surface. We uncover an essential property of the semi-Dirac systems, namely: pseudospin population with the net valley index, attainable in modified honeycomb lattices. Due to the sublattice structure, they possess a similar charge-pseudospin coupling effect and show that, due to their anisotropic Fermi surface, the accumulated nonequilibrium pseudospin population at the interface can be used for the valleytronic.

A similar model for the pseudospin-valley interplay can be realized in the superlattice of graphene patterned with a periodic potential\cite{lima2015electronic,li2021anisotropic}. It has been experimentally shown that using external potential modulations, graphene superlattice manifests pseudospin and the Fermi velocity anisotropy, which dramatically alter the quasiparticles' dynamics\cite{park2008new,park2008electron}.

\begin{acknowledgements}
This study was funded by Scientific Research Projects Coordination Unit of Istanbul University project number M2019-34733. We greatly appreciate the invaluable discussions with Mark O. Goerbig in the theory group at LPS, Orsay.
\end{acknowledgements}

\begin{widetext}
\appendix
\section{Perturbative solutions}
\subsection{zeroth correction}
\label{appx1}
Using this parametrization the stationary source term comes out to be
\begin{align}
\mathfrak{L}_{\textbf{k}\:\varepsilon_{_{\!F}}}^{(0)}(\rho)&=\frac{
(1+\boldsymbol{\sigma}\cdot{\hat{\mh}})n+\hat{\mh}\cdot\mathbf{s}+\mathbf{s}\cdot\boldsymbol{\sigma}}{\tau}.
\end{align}
If we substitute this back into Eq.~(\ref{eq:general_sol}), we find
\begin{align}\label{eq:appx_zero}
\mathfrak{g}_{\textbf{k},\varepsilon}^{(0)}&=\Lambda^{(0)}+\boldsymbol{\Gamma}^{(0)}\cdot\boldsymbol{\sigma},
\end{align}
where
\begin{subequations}\label{subzer}
\begin{align}
\Lambda^{(0)}&=n+\hat{\mh}\cdot\mathbf{s},\\
\boldsymbol{\Gamma}^{(0)}&=n\:\hat{\mh}+\frac{\zeta_-}{\tau}\;\mathbf{s}+\frac{2\zeta_3}{\tau}\;(\mathbf{s}\cdot\hat{\mh})\:\hat{\mh}+\frac{\zeta_2}{\tau}\:\hat{\mh}\times\mathbf{s},
\end{align}
\end{subequations}
where we defined $\zeta_{\pm} = \zeta_1\pm\zeta_3$.
In the clearn limit $\gamma\gg1$ the coefficients are $\zeta_1=\zeta_3\approx1/2$, $\zeta_2\approx0$. For simplifying the caluclagtions, we adopt the 4-vector structure for the pseudospin such that for the matrix components we write $\sigma^i=(I,\sigma_x,\sigma_y\sigma_z)$ and for the density components $s^i=(n, s_x,s_y,s_z)$ and $i=0,1,2,3$ indices the components of the 4-vector. The zeroth correction in the compact form becomes
\begin{align}
\mathfrak{g}_{\textbf{k},\varepsilon}^{(0)}=\sigma^ia_{ij}s^j,
\end{align}
where the coefficients can be read-off from (\ref{subzer}) as
\begin{equation*}
\begin{array}{llll}
a^{00}=1,&\qquad 							a^{01}=a^{10}=h_x,\\
a^{02}=a^{20}=h_y,& \qquad 					a^{11}=\frac{1}{2}(1+h_x^2-h_y^2),\\  
a^{12}=a^{21}=h_xh_y,&  \qquad 				a^{22}=\frac{1}{2}(1-h_x^2+h_y^2),\\
\end{array}
\end{equation*}
and $a^{i3}=a^{3i}=0$.
Now, substituting this solution back into the source term in Eq.~(\ref{eq:sors_2}) will generate the first corrected solution. Notice that the velocity operator is written as $\mathbbl{v}_i^\alpha={v}_{ij}^\alpha\sigma_j$ such that the anisotropic velocities are defined as $v_{ij}^\alpha=v_i^\alpha\:\delta_{ij}$. By restoring the dimensions, $v_x^\alpha=v_{F,x}\:\hat{k_x}^{\alpha-1}$ and $v_y^\alpha=v_{F,y}$, where $v_{F,x}=v_{F,y}=v_F$ for graphene. Then, according to Eq.~(\ref{eq:sors_2}), we can write
\begin{align}
\mathfrak{L}_{\textbf{k}z}^{(1)}\:[\mathfrak{g}_{\textbf{k},\varepsilon}^{(0)}]
&=-{v}_{ij}(\partial_i{\Gamma}_j^{(0)}+\partial_i\Lambda^{(0)}\:\sigma_j),
\end{align}
which in light of Einstein summation convention, yields the first iteration term as
\begin{align}\label{eq:appx_1}
\mathfrak{g}_{\textbf{k},\varepsilon}^{(1)}&=\Lambda^{(1)}+\boldsymbol{\Gamma}^{(1)}\cdot\boldsymbol{\sigma},
\end{align}
where
\begin{align}
\Lambda^{(1)}&=-\zeta_+{v}_{ij}\partial_i{\Gamma}_j^{(0)},\\
{\Gamma}_j^{(1)}&=-\mathcal{R}_{ij}\partial_i\Lambda^{(0)},\\
\mathcal{R}_{ij}& = \zeta_-\:{v}_{ij}+2\zeta_3\:{v}_{i\ell}\hat{{h}}_\ell\hat{{h}}_j+2\zeta_2{v}_{i\ell}h_r\:\varepsilon_{r\ell j}.
\end{align}
\iffalse
and also we defined
\begin{subequations}
\begin{align}
\mathcal{F}_{xx}&=v_x^{\alpha}(\zeta_-+2\zeta_3\:\hat{h}_x^2),\\
\mathcal{F}_{yy}&=v_y^\alpha(\zeta_-+2\zeta_3\:\hat{h}_y^2),\\
\mathcal{F}_{xy}&=2\zeta_3\:v_x^\alpha\hat{h}_x\hat{h}_y,\\
\mathcal{F}_{yx}&=2\zeta_3\:v_y^\alpha\hat{h}_x\hat{h}_y,\\
\mathcal{F}_{xz}&=-2\zeta_2\:v_x^\alpha \hat{h}_y,\\
\mathcal{F}_{yz}&=2\zeta_2\:v_y^\alpha \hat{h}_x.
\end{align}
\end{subequations}
\fi
So far, we only compute the first correction to the generalized distribution function, namely, $\mathfrak{g}_{\textbf{k},\varepsilon}^{(1)}$. According to Eqs.~(\ref{perturb}) and (\ref{perturb2}), to compute the next correction, we need the source term written as
\begin{align}
\mathfrak{L}_{\textbf{k}z}^{(1)}\:[\mathfrak{g}_{\textbf{k},\varepsilon}^{(1)}]
=&v_{ij}(\mathcal{R}_{\ell j}\:\partial_{i\ell}\Lambda^{(0)}+\zeta_+{v}_{r\ell}\:\partial_{ir}{\Gamma}_\ell^{(0)}\;\sigma_j).
\end{align}
This explicitly gives
\begin{align}
\mathfrak{g}_{\textbf{k},\varepsilon}^{(2)}&=\Lambda^{(2)}+\boldsymbol{\Gamma}^{(2)}\cdot\boldsymbol{\sigma},
\end{align}
where
\begin{subequations}
\begin{align}
\Lambda^{(2)}&=\zeta_+{v}_{ij}\mathcal{R}_{\ell j}\:\partial_{i\ell}\Lambda^{(0)},\qquad
{\Gamma}_j^{(2)}=\zeta_+{v}_{m\ell}\mathcal{R}_{ij}\:\partial_{im}{\Gamma}_\ell^{(0)}.
\end{align}
\end{subequations}

\subsection{First correction}
Using the four-vector notation, we introduced earlier and after some calculations we find the first correction as
\begin{align}
\mathfrak{g}^{(1)}&=
\sigma^i\:b_{\ell}^{ij}\:\partial_\ell s^j
\end{align}
with the coefficients $b_\ell^{ij}$ as
\begin{equation*}
\begin{array}{llll}
b_1^{00}=-\tau v_{F,x}h_x,& 							b_2^{00}=-\tau v_{F,y}h_y,\\
b_1^{01}=-\frac{\tau}{2}v_{F,x}(1+h_x^2-h_y^2),&  			b_2^{01}=-\tau v_{F,y}h_xh_y,\\
b_1^{02}=-\tau h_x h_y v_{F,x},&   						b_2^{02}=-\frac{\tau}{2} v_{F,y}(1-h_x^2+h_y^2),\\   

b_1^{10}=-\frac{\tau}{2} v_{F,x} (1+h_x^2-h_y^2),&   		b_2^{10}=-\tau v_{F,y}h_xh_y,\\
b_1^{11}=-\frac{\tau}{2} v_{F,x}h_x (1+h_x^2-h_y^2),& 		b_2^{11}=-\tau v_{F,y}h_x^2h_y,\\
b_1^{12}=-\frac{\tau}{2} v_{F,x}h_y (1+h_x^2-h_y^2),& 		b_2^{12}=-\tau v_{F,y}h_xh_y^2,\\

b_1^{20}=-\tau v_{F,x}h_xh_y,& 						b_2^{20}=-\frac{\tau}{2} v_{F,y} (1-h_x^2+h_y^2),\\
b_1^{21}=-\tau v_{F,x}h_x^2h_y,& 						b_2^{21}=-\frac{\tau}{2} v_{F,y}h_x (1-h_x^2+h_y^2),\\
b_1^{22}=-\tau v_{F,x}h_xh_y^2,&						b_2^{22}=-\frac{\tau}{2} v_{F,y}h_y (1-h_x^2+h_y^2),\\
\end{array}
\end{equation*}
and $b_i^{j3}=b_i^{3j}=0$.

\subsection{Second correction}
We similarly parametrize the second correction in the 4-pseudospin vector as
\begin{align}
\mathfrak{g}^{(2)}&=\sigma^i\:c_{\ell m}^{ij}\:\partial_{\ell m} s^j
\end{align}
where the coefficients are
\begin{equation*}
\begin{array}{llll}
c_{11}^{00}=\frac{1}{2} \left(h_x^2-h_y^2+1\right) \tau^2 v_{F,x}^2,&						c_{12}^{01}=2 h_x^2 h_y \tau^2 v_{F,x} v_{F,y},\\
c_{11}^{01}=\frac{1}{2} h_x \left(h_x^2-h_y^2+1\right) \tau^2 v_{F,x}^2,&					c_{12}^{00}=2 h_x h_y \tau^2 v_{F,x} v_{F,y},\\								 			
c_{22}^{00}=\frac{1}{2} \left(-h_x^2+h_y^2+1\right) \tau^2 v_{F,y}^2,&					c_{12}^{02}=2 h_x h_y^2 \tau^2 v_{F,x} v_{F,y},\\				
c_{22}^{01}=\frac{1}{2} h_x \left(-h_x^2+h_y^2+1\right) \tau^2 v_{F,y}^2,&					c_{22}^{10}=h_x h_y^2 \tau^2 v_{F,y}^2,\\
c_{11}^{02}=\frac{1}{2} h_y \left(h_x^2-h_y^2+1\right) \tau^2 v_{F,x}^2,&					c_{11}^{20}=h_x^2 h_y \tau^2 v_{F,x}^2,\\												
c_{22}^{12}=\frac{1}{2} h_y \left(-h_x^2+h_y^2+1\right) \tau^2 v_{F,y}^2,&					c_{22}^{11}=h_x^2 h_y^2 \tau^2 v_{F,y}^2\\			
c_{11}^{10}=\frac{1}{4} h_x \left(\left(h_x^2+1\right){}^2-h_y^4\right) \tau^2 v_{F,x}^2,&		c_{11}^{22}=h_x^2 h_y^2 \tau^2 v_{F,x}^2\\
c_{12}^{10}=-\frac{1}{2} h_y \left(-3 h_x^2+h_y^2-1\right) \tau^2 v_{F,x} v_{F,y},&			c_{11}^{11}=\frac{1}{4} \left(h_x^2-h_y^2+1\right){}^2 \tau^2 v_{F,x}^2\\
c_{12}^{12}=-\frac{1}{4} \left(h_x^4-6 h_y^2 h_x^2+h_y^4-1\right) \tau^2 v_{F,x} v_{F,y}& 		c_{22}^{02}=\frac{1}{2} h_y \left(-h_x^2+h_y^2+1\right) \tau^2 v_{F,y}^2\\
c_{12}^{11}=h_x h_y \left(h_x^2-h_y^2+1\right) \tau^2 v_{F,x} v_{F,y}&					c_{11}^{12}=\frac{1}{2} h_x h_y \left(h_x^2-h_y^2+1\right) \tau^2 v_{F,x}^2\\
c_{12}^{20}=-\frac{1}{2} h_x \left(h_x^2-3 h_y^2-1\right) \tau^2 v_{F,x} v_{F,y}&				c_{22}^{20}=\frac{1}{4} h_y \left(\left(h_y^2+1\right){}^2-h_x^4\right) \tau^2 v_{F,y}^2\\
c_{11}^{21}=\frac{1}{2} h_x h_y \left(h_x^2-h_y^2+1\right) \tau^2 v_{F,x}^2&				c_{12}^{21}=-\frac{1}{4} \left(h_x^4-6 h_y^2 h_x^2+h_y^4-1\right) \tau^2 v_{F,x} v_{F,y}\\
c_{22}^{21}=\frac{1}{2} h_x h_y \left(-h_x^2+h_y^2+1\right) \tau^2 v_{F,y}^2&				c_{12}^{22}=h_x h_y \left(-h_x^2+h_y^2+1\right) \tau^2 v_{F,x} v_{F,y}\\
c_{22}^{22}=\frac{1}{4} \left(-h_x^2+h_y^2+1\right){}^2 \tau^2 v_{F,y}^2&\\																													
\end{array}
\end{equation*}

and similarly $c_{i\ell}^{j3}=c_{i\ell}^{3j}=0$.
\section{Angular integrals}
\label{App.B}
During the momentum integrations, different combinations of momenta appear in the expressions such as
\begin{equation}
\langle{\hat{h}_x^n\hat{h}_y^m}\rangle,\quad\langle{k_x\hat{h}_x^n\hat{h}_y^m}\rangle,
\quad\text{and}\quad \langle{k_x^2\hat{h}_x^n\hat{h}_y^m}\rangle,
\end{equation}
where $n$ pertains to the longitudinal (transport) and $m$ indicates the transverse directions and are integers in both cases $\alpha=1$ and $\alpha=2$. Noting (\ref{jacobian}), the integrals are trivial to compute for graphene ($\alpha=1$) and we summarize them in the Tab.(\ref{table1}). For the semi-Dirac dispersion $\alpha=2$ the angular averaging consists of computing elliptic integrals. In this case, we assume a merged Fermi surface and Fermi energy well above the energy gap $\varepsilon\gg\Delta$, i . e., $|\eta|\ll1$. Therefore the boundaries of the integrals are defined with the angle $\theta_0=-\pi/2$. We represent the different combinations as
\begin{align}
\langle{\hat{h}_x^n\hat{h}_y^m}\rangle&=\frac{\varepsilon_{F}^{^{1/2} }}{2}\;T_{nm},\\
\langle{k_x\hat{h}_x^n\hat{h}_y^m}\rangle&=\frac{{\varepsilon_F}}{2}\;R_{nm},\\
\langle{k_x^2\hat{h}_x^n\hat{h}_y^m}\rangle&=\frac{\varepsilon_F^{^{3/2}}}{2}\;L_{nm}
\end{align}
with the associated elliptic integrals given by
\begin{align}
T_{nm}&=\int_{-\pi/2}^{\pi/2} \frac{d\theta}{2\pi} \;\frac{\cos^n\theta\sin^m \theta}{\sqrt{\cos\theta-\eta }},\\
R_{nm}&=\int_{-\pi/2}^{\pi/2} \frac{d\theta}{2\pi}\;\cos^n\theta\sin^m\theta
=\frac{1+(-1)^m}{\pi}\;{B\left( \frac{m+1}{2}, \frac{n+1}{2}\right)},\\
L_{nm}&=\int_{-\pi/2}^{\pi/2} \frac{d\theta}{2\pi} \;\sqrt{\cos\theta-\eta }\;\cos^n\theta\sin^m \theta,
\end{align}
where $B(x,y)$ is the Beta function\cite{Mathematica,*gradshteyn2014table}, and see Table (\ref{table1}) for the values of $T_{nm}$ and $L_{nm}$.

 \begin{table}[h]
	\begin{center}
		\begin{tabular}{|c|c|ccccc||cccccc|}
			\cline{3-13}
			\multicolumn{2}{c}{}&\multicolumn{5}{|c||}{\text{Semi-Dirac}\: ($\alpha=2)$}&\multicolumn{5}{c}{\text{Dirac} \;($\alpha=1$)}& \\
			\cline{2-13}
			\multicolumn{1}{c|}{}&\diagbox{n}{m}&0&1&2&3&4&0&1&2&3&4&\\
			\hline
			\multirow{5}{*}{$T_{nm}$}
			&0&0.83&0&0.55&0&0.47&1&0&$\frac{1}{2}$&0&$\frac{3}{8}$&\\
			&1&0.38&0&0.15&0&    &0&0&0&0&&\\
			&2&0.77&0&0.15& &    &$\frac{1}{2}$&0&$\frac{1}{8}$&&&\\
			&3&0.22&0&    & &    &0&0&    &&&\\
			&4&0.19& &    & &    &$\frac{3}{8}$&&    &&&\\ \hline\hline
			\multirow{5}{*}{$R_{nm}$}
			&0&0.50&0&0.25&0&0.18&0&0&0&0&0&\\
			&1&0.31&0&0.10&0&&$\frac{1}{2}$&0&$\frac{1}{8}$&0&&\\
			&2&0.25&0&0.06&&&0&0&0&&&\\
			&3&0.21&0&    &&&$\frac{3}{8}$&0&    &&&\\
			&4&0.18&&    &&&0&&    &&&\\ \hline\hline
			\multirow{5}{*}{$L_{nm}$}
			&0&0.38&0&0.15&0 &0.1&$\frac{1}{2}$&0&$\frac{1}{8}$&0&$\frac{1}{16}$&\\
			&1&0.27&0&0.07&0 &    &0&0&0&0&&\\
			&2&0.22&0&0.05& &    &$\frac{3}{8}$&0&$\frac{1}{16}$&&&\\
			&3&0.19&0&    & &    &0&0&    &&&\\
			&4&0.17& &    & &    &$\frac{5}{16}$&&    &&&\\ \hline
		\end{tabular}
		\vspace{6pt}
		\caption{Table demonstrates the values of $T_{nm}$, $R_{nm}$ and $L_{nm}$ for $\alpha=1$ and $\alpha=2$.}
		\label{table1}
	\end{center}
\end{table}

Applying the momentum integration for the semi-Dirac system and neglecting mixed derivatives $\partial_{xy}$ results in the drift-diffusion equation 

\vspace{.5in}
\resizebox{\textwidth}{!}{
$
\begin{array}{llll}
\tau^2v_{F,x}^2(0.28\:\partial_x^2n+0.22\:\partial_y^2n+0.25\:\partial_x^2s_x+0.13\:\partial_y^2s_x)-\tau v_{F,x}(0.38\:\partial_xn+0.28\:\partial_xs_x+0.22\:\partial_ys_y)=0.5n-0.38s_x,\\
\nonumber\\
\tau^2v_{F,x}^2(0.27\:\partial_x^2n+0.15\:\partial_y^2n+0.22\:\partial_x^2s_x+0.11\:\partial_y^2s_x)-\tau v_{F,x}(0.28\:\partial_xn+0.25\:\partial_xs_x+0.15\:\partial_ys_y)=-0.5s_x+0.38n,\\
\nonumber\\
\tau^2v_{F,x}^2(0.11\:\partial_x^2s_y+0.15\:\partial_y^2s_y)-\tau v_{F,x}(0.22\:\partial_yn+0.13\:\partial_ys_x+0.15\:\partial_xs_y)=0.78\:s_y.
\end{array}$
}
\vspace{.5in}

Now combing the first and the second differential equations and assuming diffusion along the $x$-direction result in
\begin{align}\label{eq:B8}
\tau v_F(-0.1\partial_xn-0.03 \partial_xs_x+0.01\tau v_F\partial_x^2n+0.03\tau v_F\partial_x^2s_x)-0.12 (n+s_x)=0
\end{align}
next neglecting the terms of lower order of magnetude we obtain (\ref{eq:conti_semi}).

\end{widetext}

\bibliographystyle{apsrev4-1}
\bibliography{Quad_refs}
\onecolumngrid

\end{document}